\documentclass[onecolumn,showpacs,preprintnumbers]{revtex4}
\usepackage[english]{babel}
\usepackage{graphicx}
\usepackage{dcolumn}
\usepackage{ifthen}
\usepackage{bm}

\begin{document}

\preprint{APS}

\title{ Transport theory of coupled quantum dots based on auxiliary operator method}
\author{Jung Hyun Oh}
\author{D. Ahn }
\affiliation{Institute of Quantum Information Processing and Systems, University of Seoul, Seoul 130-743, Korea}%
\author{Vladimir Bubanja}
\affiliation{Industrial Research Ltd, PO Box 31-310, Lower Hutt, New Zealand }

\date{\today}

\begin{abstract}
We formulate the theory of electron transport through coupled-quantum dots
by extending the auxiliary operator representation.
By using the generating functional technique, we derive the exact expressions for currents,
dot-occupation numbers and spin correlations, and examine them
based on the non-equilibrium Green's function method under the non-crossing approximation (NCA).
Our formulation generalizes the previous NCA approaches by allowing full occupation numbers with a finite Coulomb repulsion.
\end{abstract}


\maketitle

\section{Introduction}

Transport properties of the double quantum-dot system have been extensively studied both experimentally and theoretically.\cite{Wiel}
This artificial molecule, analogous to the two-impurity Anderson problem, provides a good platform
for examining the exciting physics of the correlated electron behavior,
such as the Kondo effect.\cite{Ng, Goldhaber,Schmid}
The double quantum-dot structure is also a fascinating subject from the point of view of possible applications
in quantum computation, where it is suggested as a basic building block, with qubits being
represented by electron spins in each quantum dot.\cite{Loss}.

The rich variety of correlated electron phenomena of a double quantum-dot system emerges from the inclusion
of the electron-electron interactions.\cite{Jones, Georges, Izumida, Aguado, Mravlje, Lee, Aguado1}
In this regard, many theoretical approaches have been developed that concentrate on a limited range of relevant parameter values,
such as infinite Coulomb repulsion, a finite interaction but under equilibrium transport conditions,
or symmetric dot occupations.
However, in order to describe the control and measurement of quantum bits in detail,\cite{Petta,Koppens}
it is necessary to develop the theory that deals with correlated electron behavior in a wide range of interaction parameters,
level occupancies in dots,
as well as to include the time-dependent perturbations.
The need for such a theory comes from the fact that in experimental studies of spin blockade in lateral coupled quantum dots,
independent tunnel barrier tuning with arbitrary dot occupations have been achieved.\cite{Johnson,Ono}
Similarly, initialization and manipulation of quantum bits requires description of sudden changes of energy levels
in the dots due to the time varying bias and gate voltages.

In this paper, we derive the expressions for the current, densities of states, dot occupancies, and spin correlations of
the double-dot system. Our approach enables the treatment of arbitrary Coulomb interactions, occupation numbers,
finite temperature, as well as time varying voltages.
To do this, we extend the auxiliary operator representation and
apply the non-equilibrium Green's functions method associated with the generating functional described
by the non-crossing approximation.\cite{Wingreen,Costi}
In order to establish the validity of our approach, we compare our results,
in a variety of situations, with the previous NCA
and exact methods like the numerical renormalization group (NRG) scheme. We find that our formulation
reproduces the previous NCA results, but deviates from the NRG method.
This is not surprising since it is well known that the NCA fails in describing the low-energy Fermi-liquid regime.
Since the vertex corrections cure the low temperature transport properties,\cite{Haule}
this work may be used for more involved further studies.

The paper is organized as follows: In Sec. II we introduce the Hamiltonian of the double-dot system,
and reformulate it in terms of the auxiliary particle operators in order to calculate the non-equilibrium
Green's function associated with the generating functional.
We derive the expressions for physical quantities via the relevant projection in the auxiliary
particle occupation number subspace.
In Sec. III, the transport properties of the double-dot system are examined by using the numerical calculations,
and are compared with the previous results.
We summarize our main results in Sec. IV. Some mathematical details are deferred to the the appendices.

\section{Calculation method}
\subsection{Hamiltonian}

We model the system, consisting of two quantum dots connected in series to the left and right electrodes, by the Hamiltonian,
\begin{equation}\label{Htotal}
{\cal H}={\cal H}_{dots}+{\cal H}_{leads}+{\cal H}_{T}.
\end{equation}
Taking the full electron-electron interaction into account,\cite{Mahan} the Hamiltonian of the coupled quantum dots is given by,
\begin{eqnarray}
{\cal H}_{dots} = \sum_{\alpha\sigma}\left\{
\epsilon_{\alpha\sigma}\!n_{\alpha\sigma}
+t_H c_{\alpha\sigma}^\dagger c_{{\bar\alpha}\sigma}
+ \frac{U_\alpha}{2} n_{\alpha\sigma} n_{\alpha{\bar\sigma}}\!
+ \frac{1}{4}\left(U_I-\frac{J}{2}\right)\sum_{\sigma'} n_{\alpha\sigma} n_{{\bar\alpha}\sigma'}
\right\}
-J\vec{S}_L\cdot\vec{S}_R,
\end{eqnarray}
where we assume that each dot $(\alpha=L,R)$ has energy levels $\epsilon_{\alpha\sigma}$
labeled with spin index $(\sigma=\uparrow,\downarrow)$, and is coherently coupled to the other $(\bar\alpha)$
with the tunneling matrix element $t_H$.
The dot number operators are given in terms of the creation (annihilation) operators
$c_{\alpha\sigma}^\dagger (c_{\alpha\sigma})$ by $n_{\alpha\sigma}= c_{\alpha\sigma}^\dagger c_{\alpha\sigma}$,
and the spin operators are given
by $\vec{S}_\alpha = \frac{1}{2} \sum_{\sigma\sigma'} \vec{\sigma}_{\sigma\sigma'}c_{\alpha\sigma}^\dagger c_{\alpha\sigma'}$,
where $\vec{\sigma}$ are the Pauli matrices. $U_\alpha$ and $U_I$ are the Coulomb interaction parameters
for electrons on the dot $\alpha$ and inter-dot, respectively, while $J$ is the exchange coupling constant.

The second and the third term in Eq. (\ref{Htotal}) describe the Hamiltonian of the leads
and tunneling between the dots and the adjacent leads;
\begin{eqnarray}
{\cal H}_{leads} &=& \sum_{k\alpha\sigma}\epsilon_{k\alpha\sigma} a^{\dagger}_{k\alpha\sigma}a_{k\alpha\sigma}, \nonumber\\
{\cal H}_{T} &=& \sum_{k\alpha\sigma}\left\{ T^\alpha_{k\sigma} a^{\dagger}_{k\alpha\sigma} c_{\alpha\sigma}
+ T^{\alpha*}_{k\sigma} c_{\alpha\sigma}^\dagger a_{k\alpha\sigma} \right\},
\end{eqnarray}
where $a_{k\alpha\sigma}^\dagger ~(a_{k\alpha\sigma})$ creates (annihilates) an electron at the lead $\alpha$,
and the constants $T^\alpha_{k\sigma}$ provide the coupling strength between the dot and the adjacent lead.
We assume that the energy levels of each dot are controlled independently by the nearby gate electrodes\cite{Wiel}
and the chemical potential $\mu_\alpha$ at the lead $\alpha$ is
adjusted by applied voltage difference $\Delta\mu$ to be
\begin{eqnarray}
\mu_L = \mu_0 -\Delta \mu/2,\nonumber\\
\mu_R = \mu_0 +\Delta \mu/2,
\end{eqnarray}
with respect to the equilibrium chemical potential $\mu_0$.

\begin{table}
\caption{
Schematic representation of the eigenstates for two coupled quantum dots with a coupling strength $t_H$. Here,
energy splitting for one- and three-particle states are given by
$\Delta_{1\sigma}\!=\!\sqrt{ (\epsilon_{R\sigma}\!-\!\epsilon_{L\sigma})^2/4\!+\!t_H^2}$
and
$\Delta_{3\sigma}=\sqrt{ (\epsilon_{L\sigma}\!+\!U_L\!-\!\epsilon_{R\sigma}\!-\!U_R)^2/4\!+\!t_H^2}$, respectively.
Energies of two-particle singlet states are solutions of a cubic equation,
$\Delta\epsilon^3\!+\!(U_L\!+\!U_R\!-\!U_I\!-\!J)(\Delta\epsilon^2\!-\![E_L\!-\!E_R]^2)
\!-\!([E_L\!-\!E_R]^2\!+\!16t_H^2)\Delta\epsilon=0$ and their eigenstates are determined by
$u_k=(\Delta\epsilon_k\!+\!E_L\!-\!E_R)w_k/\sqrt{8}t_H$ and
$v_k=(\Delta\epsilon_k\!+\!U_L\!+\!U_R\!-\!U_I\!-\!J)u_k/\sqrt{8}t_H-w_k$
under a normalization of $u_k^2+v_k^2+w_k^2=1$.
We also abbreviate $E_{\alpha}=\epsilon_{\alpha\uparrow}\!+\!\epsilon_{\alpha\downarrow}\!+\!U_{\alpha}$.
}
\begin{ruledtabular}
\begin{tabular}{lll}
0-particle state & $\mid\! m\!=\!0\rangle \equiv \mid e\rangle$ & $\epsilon_0 = 0$  \\
1-particle state  & $\mid\! m\!=\!1,2\rangle =(\cos\phi_\sigma c^\dagger_{L\sigma}\!-\!\sin\phi_\sigma c^\dagger_{R\sigma})\mid e\rangle$
& $\epsilon_{1,2} = (\epsilon_{L\sigma}\!+\!\epsilon_{R\sigma})/2\!-\!\Delta_{1\sigma}$  \\
                   & $\mid\! m\!=\!3,4\rangle =(\sin\phi_\sigma c^\dagger_{L\sigma}\!+\!\cos\phi_\sigma c^\dagger_{R\sigma})\mid e\rangle$
& $\epsilon_{3,4} = (\epsilon_{L\sigma}\!+\!\epsilon_{R\sigma})/2\!+\!\Delta_{1\sigma}$  \\
&with $\tan 2\phi_\sigma \!=\!2t_H/(\epsilon_{R\sigma}\!-\!\epsilon_{L\sigma})$  & \\
2-particle state & $\mid\! m\!=\!5\rangle = c_{L\uparrow}^\dagger c_{R\uparrow}^\dagger \mid e\rangle$ &
$\epsilon_5 = \epsilon_{L\uparrow}\!+\!\epsilon_{R\uparrow}\!+\!(U_I\!-\!J)/2$  \\
 & $\mid\! m\!=\!6\rangle = c_{L\downarrow}^\dagger c_{R\downarrow}^\dagger \mid e\rangle$
& $\epsilon_6 = \epsilon_{L\downarrow}\!+\!\epsilon_{R\downarrow}\!+\!(U_I\!-\!J)/2 $\\
& $\mid\! m\!=\!7\rangle = \frac{1}{\sqrt{2}}(c_{L\uparrow}^\dagger c_{R\downarrow}^\dagger\!+\!c_{L\downarrow}^\dagger
c_{R\uparrow}^\dagger)\mid e\rangle$ & $\epsilon_7 = (\epsilon_5\!+\!\epsilon_6)/2$  \\
& $\mid\! m\!=\!k\rangle = (u_k\hat{S_g}\!+\!
v_k c_{L\uparrow}^\dagger c_{L\downarrow}^\dagger\!+\!w_kc_{R\uparrow}^\dagger c_{R\downarrow}^\dagger)\mid\! e\rangle$ &
$\epsilon_{k} = (E_L\!+\!E_R)/2\!+\!\Delta\epsilon_k/2$,\\
&where $\hat{S_g}= \frac{1}{\sqrt{2}}(c_{L\uparrow}^\dagger c_{R\downarrow}^\dagger\!-\!  c_{L\downarrow}^\dagger c_{R\uparrow}^\dagger), $
& $k=8,9,10$  \\
3-particle state  & $\mid\! m\!=\!11,12\rangle =(\cos\theta_\sigma c_{L\sigma}\!-\!\sin\theta_\sigma c_{R\sigma})\mid f\rangle$
& $\epsilon_{11,12} =(\epsilon_{15}\!+\!\epsilon_{L{\bar\sigma}}\!+\!\epsilon_{R{\bar\sigma}})/2 \!-\!\Delta_{3\sigma}$  \\
& $\mid\! m\!=\!13,14\rangle =( \sin\theta_\sigma c_{L\sigma}\!+\! \cos\theta_\sigma c_{R\sigma})\mid f\rangle$
& $\epsilon_{13,14} =(\epsilon_{15}\!+\!\epsilon_{L{\bar\sigma}}\!+\!\epsilon_{R{\bar\sigma}})/2\!+\!\Delta_{3\sigma}$  \\
&with $\tan 2\theta_\sigma \!=\!2t_H/(\epsilon_{R\sigma}\!+\!U_R\!-\!\epsilon_{L\sigma}\!-\!U_L)$  & \\
4-particle state & $\mid\! m\!=\!15\rangle = c^\dagger_{L\uparrow} c^\dagger_{L\downarrow} c^\dagger_{R\uparrow} c^\dagger_{R\downarrow} \mid e\rangle \equiv \mid\! f\rangle$
 & $\epsilon_{15} = E_L\!+\!E_R\!+\!2U_I\!-\!J$  \\
\end{tabular}
\end{ruledtabular}
\label{table1}
\end{table}

In order to take into account all the possible occupancies of the double-dot system (ranging from zero to four),
we extend the idea of the
auxiliary particle representation \cite{NZA,Haule} and introduce the
auxiliary operators $d_m^\dagger (d_m)$ as,
\begin{eqnarray}
c_{\alpha\sigma}^\dagger = \sum_{mm'}\xi^{\alpha\sigma}_{mm'} d_m^\dagger d_{m'}.
\label{Convert}
\end{eqnarray}
Here, $d_m^\dagger\mid\!{\rm\!vac}\rangle$ is chosen as the $m-$th
basis vector diagonalizing the isolated coupled-quantum dots (or molecular states)
as specified in Table \ref{table1}. The auxiliary operators satisfy the commutation relations
$[d_{m'},d_m^\dagger]_{\varsigma}=\delta_{m,m'}$ and
$[d_{m'},d_m]_{\varsigma}=[d_{m'}^\dagger,d_m^\dagger]_{\varsigma}=0$,
where odd (even)-numbered states are assumed to be fermionic
(bosonic) and  $\varsigma\!=\!+$ if both $m$ and $m^\prime$ denote fermionic states, otherwise
$\varsigma\!=\!-$.
The overlap matrix $\xi^{\alpha\sigma}_{mm'}=\langle
m\mid\!  c_{\alpha\sigma}^\dagger\!\mid m'\rangle$
has a non-zero value for the combination of boson-fermion or fermion-boson operators to
ensure original commutation relation of
$[c_{\alpha\sigma},c^\dagger_{\alpha'\sigma'}]=\delta_{\alpha,\alpha'}\delta_{\sigma,\sigma'}$
under the constraint of $Q=\sum_m d_m^\dagger d_m=1$. The proof is given in Appendix \ref{dcomm}.
In terms of the auxiliary operators the Hamiltonian of the coupled dots is given by,
\begin{eqnarray}
{\cal H}_{dots} &=& \sum_{m=0}^{15} \epsilon_m d^\dagger_m d_m+{\cal H}_{int}(\{d_{m^\prime},d_m^\dagger\}).
\end{eqnarray}
The term ${\cal H}_{int}$ represents the interaction between auxiliary particles (applicable for $Q\geq 2$) and we omit it hereafter since it doesn't affect our final results.

\subsection{Generating functional}

For the ease in evaluating the expectation value of any operator ${\cal O}$,
we introduce a Lagrange multiplier $\lambda$ associated with the auxiliary particle number $Q$ as,
\begin{eqnarray}
{\cal H} \rightarrow {\cal H}+\lambda Q-\sum_\alpha \mu_\alpha n_\alpha^{leads},
\end{eqnarray}
with $n_\alpha^{leads}=\sum_{k\sigma} a^\dagger_{k\alpha\sigma} a_{k\alpha\sigma}$.
Then, the system becomes the grand canonical ensemble with respect to the auxiliary particle number $Q$,
{\it i.e.}, $Q$ is now unconstrained.

With the grand canonical ensemble,
we define a generating functional ${\cal W}=-\ln{\cal Z}$ as an extension of the Gibbs free energy.
Here, the generalized partition function ${\cal Z}$, in terms of the coherent path integral representation is given by,\cite{Negele,Oh}
\begin{eqnarray}
{\cal Z} = \oint {\cal D}[c_{\alpha\sigma}^*,c_{\alpha\sigma},
a^*_{k\alpha\sigma},a_{k\alpha\sigma}] e^{-S/i\hbar},
\label{partitionFn}
\end{eqnarray}
with the action represented on a closed time contour as,
\begin{eqnarray}
S &=& \oint d\tau\Big [\sum_{m}d^*_{m}(\tau)\big(i\hbar\partial\tau-\epsilon_{m}\!-\!\lambda \big)d_{m}(\tau)
+\sum_{k\alpha\sigma}a^*_{k\alpha\sigma}(\tau)\big(i\hbar\partial\tau -\epsilon_{k\alpha\sigma}
+\mu_\alpha\big )a_{k\alpha\sigma}(\tau)
\nonumber\\
&-&\sum_{k\alpha\sigma}\big\{  T^\alpha_{k\sigma} a^*_{k\alpha\sigma}(\tau)c_{\alpha\sigma}(\tau)
+T^{\alpha *}_{k\sigma} c^*_{\alpha\sigma}(\tau)a_{k\alpha\sigma}(\tau) \big\}\Big ].
\end{eqnarray}
We note that the Fermi (Bose) particle operators are now replaced by the corresponding
Grassman (complex) variables
$\{ a_{\alpha\sigma}^*,a_{\alpha\sigma},d_m^*,d_m \}$.

From the unconstrained generating functional ${\cal W}$,
the expectation value in the $Q=1$ ensemble can be calculated easily by noting that
the operator $Q$ commutes with the total Hamiltonian, and $Q$ is thus a good quantum number.
This fact enables us to expand the partition function in powers of $\zeta = e^{-\lambda\beta}$. The partition function belonging to the $Q=1$ subspace can be obtained by differentiating with respect to $\zeta$, {\it i.e.},
${\cal Z}_{Q=1} = \lim_{\lambda\rightarrow\infty} \frac{\partial}{\partial\zeta} {\cal Z}$.
Based on this relation, we can evaluate the expectation
value  of ${\cal O}$ by taking a functional derivative of ${\cal Z}$
with respect to its conjugate variable $\eta$ as,
\begin{eqnarray}
\langle {\cal O}\rangle_C
=-\frac{1}{{\cal Z}_{Q=1}}\frac{\delta}{\delta \eta}{\cal Z}_{Q=1}
=\lim_{\lambda\rightarrow\infty}\left[ \langle{\cal O}\rangle_{GC}+
\frac{(\partial/\partial\zeta)\langle{\cal O}\rangle_{GC}}{e^{\beta\lambda}\langle Q\rangle_{GC} }
\right].
\label{aveC}
\end{eqnarray}
where $\langle{\cal O}\rangle_C$ denotes the average over the canonical ensemble, {\it i.e.,} over the subspace $Q=1$ while
$\langle{\cal O}\rangle_{GC}\equiv\delta {\cal W}/\delta \eta$ is the average over the grand canonical ensemble.
When the canonical expectation value of the operator ${\cal O}$ has a zero expectation value
in the $Q=0$ subspace, the above relation is further simplified to,
\begin{eqnarray}
\langle {\cal O}\rangle_C
=\lim_{\lambda\rightarrow\infty} \frac{
\langle {\cal O}\rangle_{GC}
}{\langle Q\rangle_{GC}}
\label{aveGC}
\end{eqnarray}
As seen in the following section, since the expectation values of interest have a zero expectation value
in the $Q=0$ subspace we hereafter focus on the average over the grand canonical ensemble based on Eq. (\ref{aveGC}).

The partition function in the grand canonical ensemble is calculated following the standard series
expansion procedure.\cite{Oh} Firstly, we integrate the action over Grassman variables
$\{ a_{k\alpha\sigma}^*,a_{k\alpha\sigma}\}$
and obtain
\begin{eqnarray}
S &=& \oint d\tau\sum_{m}d^*_{m}(\tau)\big(i\hbar\partial\tau-\epsilon_{m}\!-\!\lambda \big)d_{m}(\tau)
\!-\!\oint d\tau d\tau^\prime \mid T^\alpha_{k\sigma}\mid^2 c_{\alpha\sigma}^*(\tau)
g_{k\alpha\sigma}(\tau,\tau^\prime)c_{\alpha\sigma}(\tau^\prime).
\end{eqnarray}
Then we express the variables $\{ c_{\alpha\sigma}^*,c_{\alpha\sigma}\}$ in Eq. (\ref{Convert})
in terms of their functional derivatives as,
\begin{eqnarray}
c_{\alpha\sigma}^*(\tau) &=& \sum_{mm^\prime} \xi^{\alpha\sigma}_{mm^\prime} \varsigma_m
\frac{\delta}{\delta\eta_m(\tau)}\frac{\delta}{\delta\eta^*_{m^\prime}(\tau)} \nonumber\\
c_{\alpha\sigma}(\tau) &=& \sum_{mm^\prime} \xi^{\alpha\sigma *}_{mm^\prime} \varsigma_{m^\prime}
\frac{\delta}{\delta\eta_{m^\prime}(\tau)}\frac{\delta}{\delta\eta_m^*(\tau)}
\end{eqnarray}
where $\eta_m$ and $\eta_m^*$ are their corresponding conjugate variables.
In that way the partition function is further rewritten as,
\begin{eqnarray}
{\cal Z} = {\cal Z}^0 \exp\left\{\frac{1}{i\hbar}\sum_{\alpha\sigma}\oint d\tau d\tau^\prime c_{\alpha\sigma}^*(\tau)
g_{\alpha\sigma}(\tau,\tau^\prime)
c_{\alpha\sigma}(\tau^\prime) \right\}
e^{i\hbar \sum_m \oint d\tau d\tau^\prime \eta_m^* g_m(\tau,\tau^\prime)\eta_m(\tau^\prime)}
\label{Znaive}
\end{eqnarray}
where ${\cal Z}_0$ is the unperturbed partition function and
$g_{\alpha\sigma}(\tau,\tau') \equiv \sum_k\mid\!T^\alpha_{k\sigma}\!\mid^2
g_{k\alpha\sigma}(\tau,\tau')$.
Here, the lower case $g$'s are the unperturbed Green's functions of the lead-electrons and auxiliary particles;
for instance, retarded, Keldysh, and advanced components of the Green's functions of the $m$-th auxiliary particle
are given by
\begin{eqnarray}
g^R_m(t,t') &=& \frac{1}{i\hbar}\theta(t-t')\langle[d_m(t),d_m^\dagger(t')]_{-\varsigma_m}\rangle^0_{GC}
=\frac{1}{i\hbar}\theta(t-t') e^{(\epsilon_m+\lambda)(t-t')/i\hbar}\nonumber\\
g^K_m(t,t') &=& \frac{1}{i\hbar}\langle[d_m(t),d_m^\dagger(t')]_{\varsigma_m}\rangle^0_{GC}
=
\frac{1}{i\hbar}
\left[\tanh\frac{\beta(\epsilon_m+\lambda)}{2} \right]^{-\varsigma_m}
e^{(\epsilon_m+\lambda)(t-t')/i\hbar} \nonumber\\
g^A_m(t,t') &=& g^{R*}_m(t',t)
\label{g0}
\end{eqnarray}
respectively, where the superscript '$0$' denotes the average in the case of the dots being decoupled from the leads, and $\varsigma_m$ is $-1(+1)$
if the particle $m$ is a fermion (boson).
To simplify the expressions, it is sometimes convenient to use greater, $g^>=(g^K\!+\!g^R\!-\!g^A)/2$, lesser, $g^<=(g^K\!-\!g^R\!+\!g^A)/2$,
and correlated, $g^C=\!g^R\!-\!g^A$, Green's functions interchangeably.

Next, we expand the exponential function in Eq. (\ref{Znaive}) in power series
and obtain the partition function by collecting all the connected diagrams;
\begin{eqnarray}
{\cal Z} = {\cal Z}^0 e^{-\{ {\cal W}^{(1)} +{\cal W}^{(2)} +{\cal W}^{(3)} +\cdots\}}.
\end{eqnarray}
Here, ${\cal W}^{(n)}$ is the collection of the $\mid T^\alpha_{k\sigma}\mid^n$-order diagrams.
For instance, ${\cal W}^{(1)}$ and ${\cal W}^{(2)}$ look like,
\begin{eqnarray}
{\cal W}^{(1)} &=& -i\hbar\sum_{m} \varsigma_m
\begin{array}{l}
 \includegraphics[width=8ex]{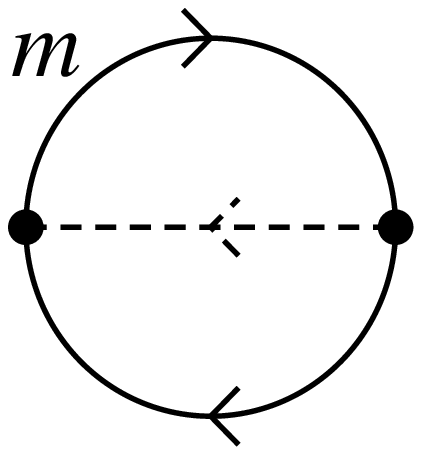} ,
\end{array}\nonumber\\
{\cal W}^{(2)} &=& -
\frac{ i\hbar }{2}\sum_{m} \varsigma_m
\left[
\begin{array}{l}
 \includegraphics[width=8ex]{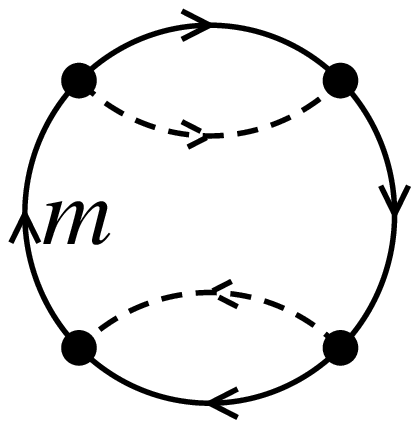}
\end{array}
+2
\begin{array}{l}
 \includegraphics[width=8ex]{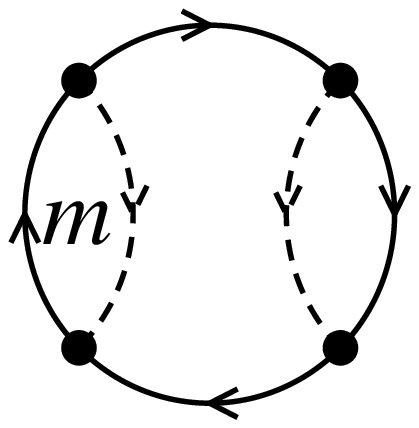}
\end{array}
+2
\begin{array}{l}
 \includegraphics[width=8ex]{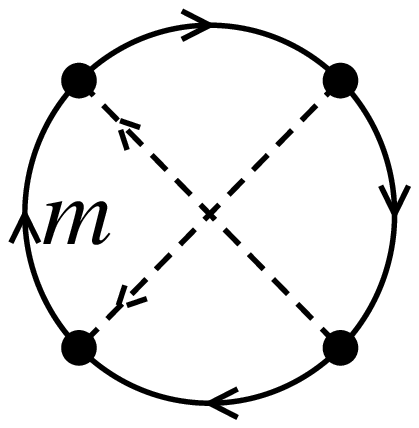}
\end{array}
+
\begin{array}{l}
 \includegraphics[width=8ex]{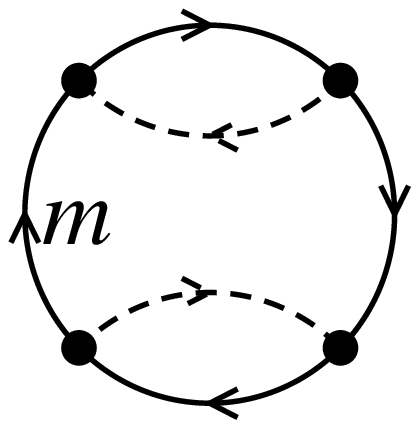}
\end{array}
\right],
\label{GF}
\end{eqnarray}
where the solid (dotted) lines denote the unperturbed Green's functions $g_m$ $(g_{\alpha\sigma})$.
Here, large dots indicate the times at which the tunneling events occur and
the overlap matrix is assumed to be multiplied as,
\begin{eqnarray}
\begin{array}{l}
 \includegraphics[width=10ex]{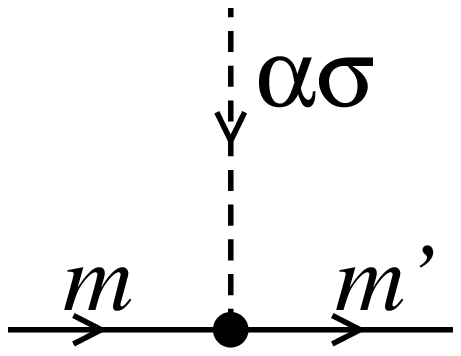}
\end{array}
\times \xi^{\alpha\sigma*}_{m'm}, ~~~~~~~~~~
\begin{array}{l}
 \includegraphics[width=10ex]{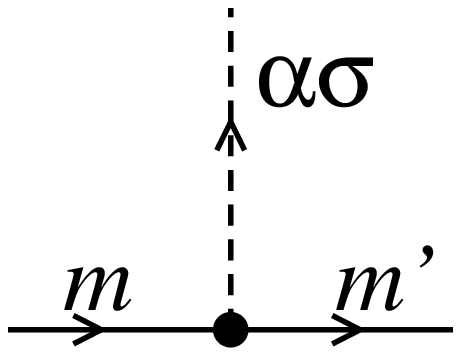}
\end{array}
\times \xi^{\alpha\sigma}_{mm'}.
\end{eqnarray}
Finally, the calculation of the generating functional ${\cal W}=-\ln{\cal Z}$ is done in a straightforward
fashion via the Luttinger-Ward functional $\Phi$ and repeated terms;\cite{Luttinger}
\begin{eqnarray}
{\cal W} &=& -{\rm ln}{\cal Z}^{(0)}+{\cal W}^{(1)} +{\cal W}^{(2)} +\cdots \nonumber\\
         &=& \Phi+\sum_p \varsigma_p{\rm Tr}\left[ \ln G_p^{-1}+\Sigma_p G_p \right], ~~~~p=\alpha\sigma~{\rm and}~m.
\end{eqnarray}
Here, the Luttinger-Ward functional $\Phi$ is the sum of all the closed skeleton diagrams
with a non-interacting Green's functions $(g_p)$ replaced by the full Green's functions $(G_p)$.

Up to now, the generating functional of the coupled-dot system has been derived without any approximations,
and thus the associated Green's functions give the exact expressions for the physical quantities as shown in the Appendices
\ref{PQt} and \ref{Icon}.
In the next section we describe the approximation of the Luttinger-Ward functional $\Phi$,
and present the expressions for the physical quantities  in the static case.

\subsection{Non-crossing approximation and projection to $Q=1$ }

Hereafter we employ the non-crossing approximation, that is, we confine our attention to the first skeleton diagram
originated from ${\cal W}^{(1)}$, and approximate the Luttinger-Ward functional by,
\begin{eqnarray}
\Phi &=& -i\hbar\sum_{m} \varsigma_m
\begin{array}{l}
 \includegraphics[width=8ex]{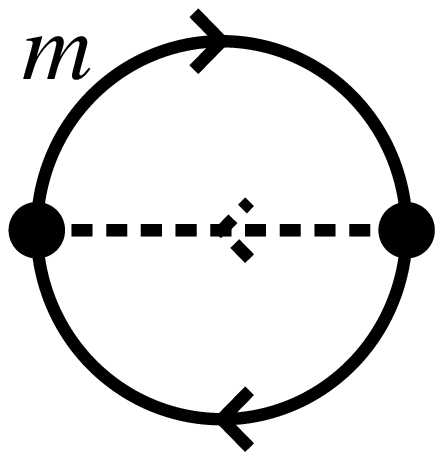}
\end{array}
\nonumber\\
&=&
-i\hbar\sum_{mm'\alpha\sigma} \varsigma_m \mid\! \xi_{m,m'}^{\alpha\sigma}\!\mid^2 \oint d\tau d\tau'
G_{m}(\tau',\tau) G_{\alpha\sigma}(\tau,\tau') G_{m'}(\tau,\tau')
\label{PhiNCA}
\end{eqnarray}
where the thick lines represent the full Green's functions of the particles
instead of unperturbed ones (thin lines) in Eq. (\ref{GF}).

Then, since the generating function $\cal W$ is stationary with respect to $G_p$, namely $\delta{\cal W}/\delta G_p=0$,
the self-energies can be obtained from,
\begin{eqnarray}
\Sigma_p(\tau,\tau')=-\varsigma_p \frac{\delta\Phi}{\delta G_p(\tau',\tau)}.
\end{eqnarray}
Using the NCA functional Eq. (\ref{PhiNCA}), this gives,
\begin{eqnarray}
\Sigma_{\alpha\sigma}(\tau,\tau') &=&
-i\hbar\sum_{mm'} \varsigma_m \mid\! \xi_{m,m'}^{\alpha\sigma}\!\mid^2
G_{m}(\tau,\tau')  G_{m'}(\tau',\tau), \nonumber \\
\Sigma_{m}(\tau,\tau') &=&
i\hbar\sum_{m'\alpha\sigma} \Big [ \mid\! \xi_{m,m'}^{\alpha\sigma}\!\mid^2 G_{\alpha\sigma}(\tau,\tau')
 -\mid\! \xi_{m',m}^{\alpha\sigma}\!\mid^2 G_{\alpha\sigma}(\tau',\tau) \Big ] G_{m'}(\tau,\tau'),
\end{eqnarray}
for the self-energies of electrons in the leads and auxiliary particles.
As functions of real time arguments, the above expressions can be rewritten as,
\begin{eqnarray}
\Sigma^R_{\alpha\sigma}(t,t^\prime;\lambda) &=&
-i\hbar\sum_{mm'} \varsigma_m \mid\! \xi_{m,m'}^{\alpha\sigma}\!\mid^2
\left[ G^R_{m}(t,t^\prime;\lambda)  G^<_{m'}(t^\prime,t;\lambda)+
G^<_{m}(t,t^\prime;\lambda)  G^A_{m'}(t^\prime,t;\lambda) \right],\nonumber\\
\Sigma^K_{\alpha\sigma}(t,t^\prime;\lambda) &=&
-i\hbar\sum_{mm'} \varsigma_m \mid\! \xi_{m,m'}^{\alpha\sigma}\!\mid^2
\left[ G^<_{m}(t,t^\prime;\lambda)  G^C_{m'}(t^\prime,t;\lambda)
+G^C_{m}(t,t^\prime;\lambda)  G^<_{m'}(t^\prime,t;\lambda)
 \right]
\label{Sigma0E}
\end{eqnarray}
for the electrons in the leads, and
\begin{eqnarray}
\Sigma^R_{m}(t,t';\lambda) &=&
i\hbar\sum_{m'\alpha\sigma}
\left [
 \mid\! \xi_{m,m'}^{\alpha\sigma}\!\mid^2 G^>_{\alpha\sigma}(t,t^\prime;\lambda)
-\mid\! \xi_{m',m}^{\alpha\sigma}\!\mid^2 G^<_{\alpha\sigma}(t^\prime,t;\lambda) \right] G^R_{m'}(t,t';\lambda),
\nonumber\\
\Sigma^<_{m}(t,t';\lambda) &=&
i\hbar\sum_{m'\alpha\sigma}
\left [
 \mid\! \xi_{m,m'}^{\alpha\sigma}\!\mid^2 G^<_{\alpha\sigma}(t,t^\prime;\lambda)
-\mid\! \xi_{m',m}^{\alpha\sigma}\!\mid^2 G^>_{\alpha\sigma}(t^\prime,t;\lambda) \right] G^<_{m'}(t,t';\lambda).
\label{Sigma0A}
\end{eqnarray}
for the auxiliary particles.

Next we project the self-energy to the $Q=1$ ensemble (details can be found
elsewhere\cite{Wingreen,Costi,Coleman}).
For this we exploit two facts about the lesser and greater components of the Green's functions for the auxiliary particles.
Firstly, it is important to note that self-energies of electrons in the leads
depend on lesser components of Green function $G^<_m$.
This makes the projection to $Q=1$ subspace easy because
\begin{eqnarray}
G_m^{<}(t,t^\prime)= \frac{\varsigma_m}{i\hbar} \langle d_m(t^\prime)^\dagger d_m(t)\rangle_{GC}
={\cal O}(e^{-\beta\lambda})
\end{eqnarray}
means a zero expectation value in the $Q=0$ subspace, and thus one can use Eq. (\ref{aveGC}) in evaluating
observables.
Secondly, the lesser and greater components of the Green's functions for the auxiliary particles are given by,
\begin{eqnarray}
G_m^{<,>}(t,t') = \int dt_1 dt_2 G_m^{R}(t,t_1)
\Sigma_m^{<,>}(t_1,t_2) G_m^{A}(t_2,t')
\label{Gmless}
\end{eqnarray}
without dependence on $g_m^{<,>}$ due to the loss of memory.\cite{Wingreen,Mahan}

Before the projection to the $Q=1$ subspace, we eliminate the $\lambda$-dependence in $g_m^{R,A}$.
Since  $\lambda$ is related to $g_m^{R,A}(t,t')$ only through the factor $e^{\lambda(t-t')/i\hbar}$ in
 Eq. (\ref{g0}), the elimination of $\lambda$ such as
$g_m(t,t';\lambda)\rightarrow g_m(t,t')\equiv g_m(t,t';\lambda\!=\!0)$ results in
the modified forms of self-energies in Eqs. (\ref{Sigma0E}) and (\ref{Sigma0A}) like,
$\Sigma_p(t,t';\lambda) \rightarrow \Sigma_p(t,t';\lambda)e^{-\lambda(t-t')/i\hbar}$.

By using  Eq. (\ref{aveGC}) and taking  the projection of $\lambda\rightarrow \infty$,
the projected self-energies of Eq. (\ref{Sigma0E}) become,
\begin{eqnarray}
\Sigma^R_{\alpha\sigma}(t,t^\prime) &=&
-i\hbar\sum_{mm'} \varsigma_m \mid\! \xi_{m,m'}^{\alpha\sigma}\!\mid^2
\left[ G^R_{m}(t,t^\prime)  G^<_{m'}(t^\prime,t)+
G^<_{m}(t,t^\prime)  G^A_{m'}(t^\prime,t) \right],\nonumber\\
\Sigma^K_{\alpha\sigma}(t,t^\prime) &=&
-i\hbar\sum_{mm'} \varsigma_m \mid\! \xi_{m,m'}^{\alpha\sigma}\!\mid^2
\left[ G^<_{m}(t,t^\prime)  G^C_{m'}(t^\prime,t)
+G^C_{m}(t,t^\prime)  G^<_{m'}(t^\prime,t)
 \right]
\label{Sas}
\end{eqnarray}
where
we use the abbreviated notation of
\begin{eqnarray}
\Sigma^{R,K}_{\alpha\sigma}(t,t')\equiv\lim_{\lambda\rightarrow\infty} e^{\lambda\beta}
\Sigma^{R,K}_{\alpha\sigma}(t,t';\lambda) e^{-\lambda(t-t')/i\hbar}
\nonumber
\end{eqnarray}
and set a $e^{\beta\lambda} \langle Q\rangle_{GC}$ term aside in Eq. (\ref{aveGC}) for a while.
Here, $G^{<}_m(t,t')$ is defined in Eq. (\ref{Gmless}) with its self-energy
$\Sigma^{<}_{m}(t,t')\equiv\lim_{\lambda\rightarrow\infty} e^{\lambda\beta}
\Sigma^{<}_{m}(t,t';\lambda) e^{-\lambda(t-t')/i\hbar}$.
Using Eq. (\ref{Sigma0A}), the self-energy is given by,
\begin{eqnarray}
\Sigma^<_{m}(t,t') =
i\hbar\sum_{m'\alpha\sigma}
\left [
 \mid\! \xi_{m,m'}^{\alpha\sigma}\!\mid^2 g^<_{\alpha\sigma}(t,t^\prime)
-\mid\! \xi_{m',m}^{\alpha\sigma}\!\mid^2 g^>_{\alpha\sigma}(t^\prime,t) \right] G^<_{m'}(t,t').
\label{Sml}
\end{eqnarray}
Whereas, the Dyson equation of $G^{R,A}_m(t,t')$ is
\begin{eqnarray}
G_{m}^{R,A}(t,t') = g_{m}^{R,A}(t,t') + \int dt_1 dt_2 g_{m}^{R,A}(t,t_1)
\Sigma_{m}^{R,A}(t_1,t_2) G_{m}^{R,A}(t_2,t')
\nonumber
\end{eqnarray}
with its self-energy defined by
$\Sigma^{R,A}_m(t,t') \equiv \lim_{\lambda\rightarrow\infty} \Sigma^{R,A}_m(t,t';\lambda) e^{-\lambda(t-t')/i\hbar}$; from
Eq. (\ref{Sigma0A}),
\begin{eqnarray}
\Sigma^R_{m}(t,t') = i\hbar\sum_{m'\alpha\sigma}
\left [
 \mid\! \xi_{m,m'}^{\alpha\sigma}\!\mid^2 g^>_{\alpha\sigma}(t,t^\prime)
-\mid\! \xi_{m',m}^{\alpha\sigma}\!\mid^2 g^<_{\alpha\sigma}(t^\prime,t) \right] G^R_{m'}(t,t').
\label{Smr}
\end{eqnarray}
During the projection, we employ the relation,
\begin{eqnarray}
G_{\alpha\sigma}^{<,>}(t,t';\lambda) &=&
g_{\alpha\sigma}^{<,>}(t,t') +
 \int dt_1 dt_2 G_{\alpha\sigma}^{R}(t,t_1)
\Sigma_{\alpha\sigma}^{<,>}(t_1,t_2;\lambda) G_{\alpha\sigma}^{A}(t_2,t')
\label{Glessas}
\end{eqnarray}
and neglect the second term, due to its ${\cal O}(e^{-\lambda\beta})$ dependence as seen from Eq. (\ref{Sigma0A}).

On the other hand, the expectation value of the operator $Q$ is given by
\begin{eqnarray}
\lim_{\lambda\rightarrow\infty} e^{\lambda\beta}\langle Q\rangle_{GC} &=&
\lim_{\lambda\rightarrow\infty} e^{\lambda\beta} i\hbar \sum_m \varsigma_m G_m^{<}(t,t;\lambda)
= i\hbar \sum_m \varsigma_m G_m^{<}(t,t)
\label{normalF}
\end{eqnarray}
where the second step can be derived in a similar way to that of Appendix \ref{PQt}.
Throughout this work, we keep
$\lim_{\lambda\rightarrow\infty} e^{\lambda\beta}\langle Q\rangle_{GC}$
 to be unity via the normalization and consequently
Eqs. (\ref{Sas})-(\ref{Sml}) are also the averaged values in the canonical ensemble.

Eqs. (\ref{Sas})-(\ref{Sml}) are the main results of this work,
which can be applied to a double-dot system at arbitrary temperature, Coulomb interaction, source-drain and gate voltage configurations, including the time-dependent problems.

\subsection{Physical quantities in static cases}

Since in the static case, the Green's functions depend only on the time interval, it becomes convenient to use the Fourier transform,
\begin{eqnarray}
G(t,t') = \frac{1}{2\pi\hbar} \int_{-\infty}^{\infty}dE e^{E(t-t')/i\hbar} G(E).
\label{Fourier}
\end{eqnarray}

By using the cut-off,
$\rho^{\alpha\sigma}_c(E)=2\pi\sum_k \mid T^\alpha_{k\sigma}\mid^2 \delta(E-\epsilon_{k\sigma})$,
the unperturbed Green's function of electrons in the lead $\alpha$, is then given in the energy space as,
\begin{eqnarray}
g^{<,>}_{\alpha\sigma}(E) &=& \pm i\hbar \rho^{\alpha\sigma}_c(E) f(\pm (E-\mu_\alpha) ),
\end{eqnarray}
where $f(E)= 1/(1+e^{\beta E})$ is the Fermi-Dirac distribution function.
By defining the spectral function $A_m$  such that $G^<_m(t,t') = -2\pi i\varsigma_m A_m(t,t')$,
the self-energies of Eqs. (\ref{Sas})-(\ref{Sml})
are rewritten as,
\begin{eqnarray}
\Sigma^R_{\alpha\sigma}(E) &=&
\sum_{mm'} \int_{-\infty}^{\infty} dE' \left[
\mid\xi_{m',m}^{\alpha\sigma}\!\mid^2 G^R_{m'}(E\!+\!E')-
\mid\xi_{m,m'}^{\alpha\sigma}\!\mid^2 G^A_{m'}(E'\!-\!E)\right]A_{m}(E'),\nonumber\\
\Sigma^K_{\alpha\sigma}(E) &=&
\sum_{mm'} \int_{-\infty}^{\infty} dE' \left[
\mid\xi_{m',m}^{\alpha\sigma}\!\mid^2 G^C_{m'}(E\!+\!E')-
\mid\xi_{m,m'}^{\alpha\sigma}\!\mid^2 G^C_{m'}(E'\!-\!E)\right]A_{m}(E') \nonumber\\
\Sigma^R_m(E) &=& \frac{i}{2\pi}
\sum_{\alpha\sigma m'} \int_{-\infty}^{\infty} dE' \left[
\mid\xi_{m,m'}^{\alpha\sigma}\!\mid^2 g^>_{\alpha\sigma}(E\!-\!E')-
\mid\xi_{m',m}^{\alpha\sigma}\!\mid^2 g^<_{\alpha\sigma}(E'\!-\!E)\right]G^R_{m'}(E').
\label{Erep}
\end{eqnarray}
Here, the spectral function $A_m(E)$ is determined from
\begin{eqnarray}
A_m(E) &=&\frac{i}{2\pi}\mid\! G^R_m(E)\!\mid^2
\sum_{\alpha\sigma m'} \int_{-\infty}^{\infty} dE' \left[
\mid\!\xi_{m',m}^{\alpha\sigma}\!\mid^2 g^>_{\alpha\sigma}(E'\!-\!E)
-\mid\!\xi_{m,m'}^{\alpha\sigma}\!\mid^2 g^<_{\alpha\sigma}(E\!-\!E')
\right]A_{m'}(E')\nonumber
\end{eqnarray}
with the normalization condition of $\sum_m \int dE A_m(E) = 1$ from Eq. (\ref{normalF}).

On the other hand, the expectation values of the physical quantities in the $Q=1$ ensemble
can be obtained by combining the results of the Appendix \ref{PQt} with Eq. (\ref{Erep}).
We summarize the results, in the energy space; for the current in the lead $\alpha$,
\begin{eqnarray}
I_{\alpha} =
\frac{q}{2\pi\hbar} \Re \sum_{\sigma}\int_{-\infty}^{\infty} dE
\left[
 g^{>}_{\alpha\sigma}(E)\Sigma_{\alpha\sigma}^{<}(E)
-g^{<}_{\alpha\sigma}(E)\Sigma_{\alpha\sigma}^{>}(E) \right],
\label{Istatic}
\end{eqnarray}
for the density of states,
\begin{eqnarray}
DOS_{\alpha\sigma}(E) = -\frac{1}{\pi}{\rm Im} {\cal G}^R_{\alpha\sigma}(E)
 = -\frac{1}{\pi}{\rm Im} \Sigma^R_{\alpha\sigma}(E),
\label{Dstatic}
\end{eqnarray}
for the occupation number,
\begin{eqnarray}
\langle n_{\alpha\sigma}\rangle_C = \sum_m \left( \frac{\partial \epsilon_m}{\partial \epsilon_{\alpha\sigma}}\right)\int_{-\infty}^{\infty}dE A_m(E),
\label{nstatic}
\end{eqnarray}
and for the spin-spin correlations,
\begin{eqnarray}
S_2 = \langle \vec{S}_L\cdot\vec{S}_R\rangle_C
=-\sum_m \frac{\partial \epsilon_m(U_I\!\rightarrow\!U_I\!+\!J/2)}{\partial J}\int_{-\infty}^{\infty}dE A_m(E).
\label{Sstatic}
\end{eqnarray}

\section{Results and Discussion}

In this section, we illustrate the numerical solutions of Eq. (\ref{Erep}),
and the resulting physical quantities of Eqs. (\ref{Istatic}) - (\ref{Sstatic}), as well as the accuracy of the present theory.

For the cut-off function we choose a Lorentzian model $\rho^{\alpha\sigma}_c(E)$
\begin{eqnarray}
\rho^{\alpha\sigma}_c(E) =\Gamma_{\alpha\sigma} \frac{W^2}{(E-\mu_\alpha)^2+W^2)}
\end{eqnarray}
with $W$ being the half width of the conduction band.

In solving the Dyson's equations with self-energies given in  Eq. (\ref{Erep}),
we use the adaptive mesh scheme where more mesh points are inserted into a high weighted region for every interaction.
The iteration is repeated until the following sum rules converge within 0.01\%;
\begin{eqnarray}
-\frac{1}{\pi}\int_{-\infty}^{\infty} {\rm Im}G^R_m(E) dE &=& 1, \nonumber\\
\sum_m \int_{-\infty}^{\infty} A_m(E) dE &=& 1. \nonumber
\end{eqnarray}
To achieve this numerical accuracy, we use about 1000 mesh points for each Green's function of an auxiliary particle.

For simplicity we consider the symmetric case, $U_L=U_R=U$, $\epsilon_{L\sigma}=\epsilon_{R\sigma}=\epsilon_d$,
$U_I=0$,
with $\Gamma_{\alpha\sigma}=\Gamma$, and $J=0$, and all the energies are measured in units of $\Gamma$ (in an experiment $\Gamma$ is typically of the order of $\mu$eV to meV).
We present results for two kinds of systems;
one is a single quantum dot (that is, we take $t_H\rightarrow\infty$) and the other is
a double quantum dot (finite $t_H$).
Although the single quantum dot case has been extensively studied,
we revisit the problem to show that our formulation indeed encompasses the previous results.


\subsection{ Single quantum dot }

We first consider a single quantum dot and examine the correlated quantum transport through it.
To do this we write $\epsilon_d \rightarrow \epsilon_d+t_H$, and take $t_H\rightarrow\infty$.
Then,  there are four low-lying states relevant to transport:
$\mid\! 0 \rangle  = \mid\! e\rangle$,
$\mid\! 1,2 \rangle = \frac{1}{\sqrt{2}} (c_{L\sigma}^\dagger-c_{R\sigma}^\dagger)\mid\!e\rangle $,
and
$\mid\! 8 \rangle  = \frac{1}{2}
(c_{L\uparrow}^\dagger-c_{R\uparrow}^\dagger)
(c_{L\downarrow}^\dagger-c_{R\downarrow}^\dagger) \mid\!e\rangle$,
while their energies are given by $\epsilon_0 =0$, $\epsilon_{1,2} = \epsilon_d$,
and $\epsilon_{8} = 2\epsilon_{d}+\frac{1}{4}(U_L+U_R+U_I+J)$, respectively.

\begin{figure}
\centering
\includegraphics[width=0.3\textwidth]{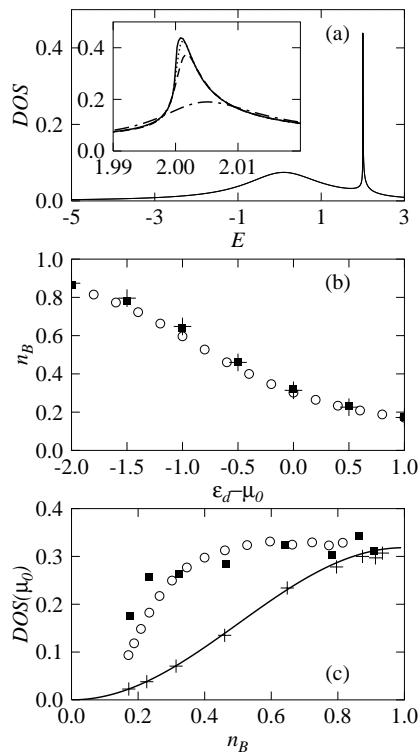}
\caption{
In (a), the equilibrium densities of states are plotted for a single quantum dot with an infinite repulsive potential.
The inset shows the blow-up around Kondo peaks for temperatures
$T=1.0\times 10^{-4}$(solid), $5.0\times 10^{-4}$(dotted), $1.0\times 10^{-3}$(dashed), and
$5.0\times 10^{-3}$(dot-dashed), respectively.
We use the parameters of $\epsilon_d=-0.5$, $\mu_0= 2.0$, $\Delta\mu=0$,
and $W =100.0$ which gives $T_K=2.8\times 10^{-3}$ in Eq. (\ref{TK}).
In (b), we compare the electronic occupation $n_B$ as a function of energy level with the previous results.
Crosses represent results from NRG, boxes from the NCA of Ref. \cite{Costi} at $T=0$
and circles from our approach
at $T=1.0\times 10^{-4}$, respectively. Since our calculation code is not available at $T=0$, we choose
sufficiently low temperature for the comparison.
In (c), the sum rule is examined together with the exact result (solid line) from Eq. (\ref{FS}).
\label{itiu}
}
\end{figure}

In the limit $U \rightarrow \infty$, the state $\mid 8 \rangle$ can be discarded further, and the present formalism
recovers the results of Ref. \cite{Wingreen}.
For this case, results of the system described by typical parameters are shown in Fig. \ref{itiu}.
In Fig. \ref{itiu}-(a), we plot the equilibrium densities of states for several temperatures,
where the broad peaks are caused by the usual transitions between levels
(in this case $\mid 0\rangle $ and $\mid 1,2\rangle$), while the sharp ones (located at $E=\mu=2.0$) are
the Kondo peaks. The later ones increase as temperature is lowered,\cite{Wingreen}
with saturation well below the Kondo temperature \cite{Haule}
\begin{eqnarray}
T_K={\rm min}\left\{\frac{U\sqrt{I}}{2\pi},\sqrt{\frac{W\hbar\Gamma}{2}} \right\} e^{-\pi/I},
\label{TK}
\end{eqnarray}
where
\begin{eqnarray}
I=\hbar\Gamma \left[ \frac{1}{\mid\epsilon_d-\mu\mid}+ \frac{1}{\epsilon_d+U-\mu\mid} \right]. \nonumber
\end{eqnarray}
From this relation, $T_K=2.8\times 10^{-3}$ is estimated in the case of Fig. \ref{itiu} while the calculated Kondo temperature,
equal to  its half width at half maximum, is $3.2\times 10^{-2}$.

The over-estimation of the Kondo temperature is a known consequence the NCA,
as well as the Kondo peak height.\cite{Wingreen}
In the figure Fig. \ref{itiu}-(b), we show the variation of the dot occupation,
$n_B=\langle n_{m=1}\rangle+\langle n_{m=2}\rangle$
as a function of the dot energy level, and in Fig. \ref{itiu}-(c),
the relation between the electronic occupation $n_B$ and the height of the density of state at $E=\mu$.
Actually, $n_B$ and the density of state are related through the Friedel sum rule:
\begin{eqnarray}
DOS(E\!=\!\mu_0)=\frac{1}{\pi\Gamma}\sin^2(\pi n_B/2).
\label{FS}
\end{eqnarray}
For the validity of our calculations, we also plot previous results of the NCA and numerical renormalization group (NRG) method
from Ref. \cite{Costi}.
In the case of the occupation, we find that our results are in good agreement with the previous results, with
a minor deviation resulting from the use of a different cut-off function.
On the other hand, in the comparison of the Friedel sum rule, a large deviation of our results are found
from those of the exact result and NRG.
As seen in results of Ref. \cite{Costi}, the previous NCA calculation also show nearly the same deviation.
This fact leads us to the over-estimated Kondo peak with the NCA in the wider range of the occupation.

\begin{figure}
\centering
\includegraphics[width=0.3\textwidth]{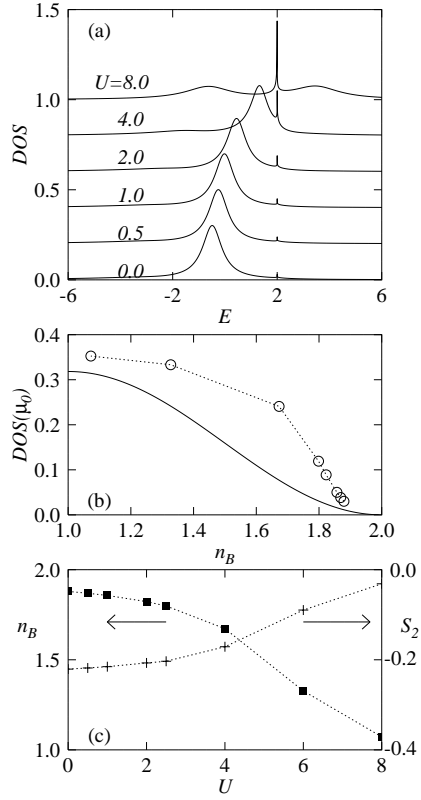}
\caption{\label{sqdk}
In (a), we show the equilibrium density of states
for the single quantum dot with the same parameters as Fig. \ref{itiu}, but
with finite potential $U$. In (b), the density of states at the chemical potential is plotted as a function of the occupation $n_B$,
where the solid line is  the Friedel sum rule of Eq. (\ref{FS}) and circles is the present result, respectively.
In (c), we plot the variation of the occupation $n_B$ and spin correlation $S_2$ with respect to the potential $U$.
Dotted lines are the guide for eyes.
}
\end{figure}

When the Coulomb interaction is finite,
all of the four states $\mid\! 0 \rangle$, $\mid\! 1 \rangle$, $\mid\! 2 \rangle$, and $\mid\! 8 \rangle$
take part in electron transport.
For the same quantum dot of Fig. \ref{itiu}, but with a finite potential $U$,
we show the DOS as a function of $E$ in Fig. (\ref{sqdk})-(a).
As the Coulomb potential decreases, the Kondo peak is lowered because the electron correlation is unimportant.
This is also predicted by the Friedel sum rule of Eq. (\ref{FS}).
Since the two-particle state $\mid\! 8 \rangle$ becomes energetically favorable with smaller potential $U$,
the electron occupation increases up to two.
In Fig. \ref{sqdk}-(b), we plot the height of the Kondo peak as a function of occupation $n_B$.
It is found that the present result (circles) exhibits the same decreasing behavior with larger occupation as the Friedel sum rule,
however still shows the over-estimation of the Kondo peak.
In Fig. \ref{sqdk}-(c), the spin correlation is shown as a function of the Coulomb potential $U$
and is compared with the occupation $n_B$.
Since the spin correlation originates from only a two-particle state,
results in the figures are proportional to the occupancy of $\mid\! 8\rangle$ auxiliary particle.
Thus, one can see that as the occupation $n_B$ approaches two, the spin correlation becomes
$S_2=\langle 8\!\mid \vec{S}_L\cdot \vec{S}_R\mid\! 8\rangle = -3/8$.

\begin{figure}
\centering
\includegraphics[width=0.3\textwidth]{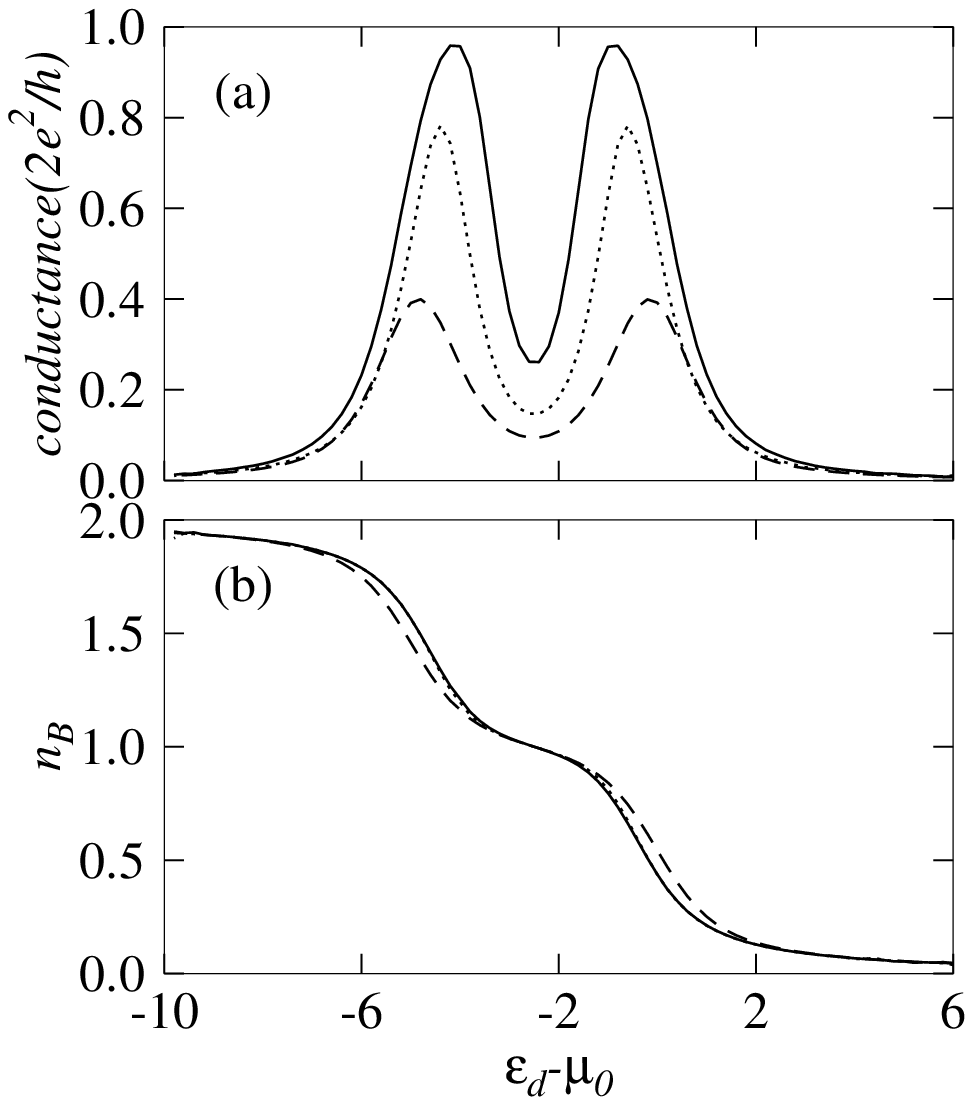}
\caption{
We plot the linear conductance (in unit of $2e^2/h$) as a function of single-particle energy in $(a)$
and corresponding electronic occupation (b), for a finite Coulomb potential $U=10.0$
at temperatures $T=0.003$(solid), $0.03$(dotted), and $0.3$(dashed), respectively.
We use $W=50.0$ and $\mu_0=4.0$. The linear conductance is calculated with a finite potential difference
$\Delta\mu=0.01$ through $I_\alpha/\Delta\mu$.
}
\label{condfU}
\end{figure}

Fig. \ref{condfU} shows the linear conductance (a), and the electronic occupation $n_B$ (b)
as a function of the single-particle energy for a finite Coulomb potential $U=10.0$.
For this calculation, we apply a small voltage between the left and right leads  of $\Delta\mu=0.01$ and
the conductance is calculated as the current at a lead divided by $\Delta\mu$.
As the temperature is lowered, the conductance increases and approaches $2e^2/h$, which is
in accordance with the experimental results reflecting the Kondo effect.\cite{Goldhaber}
On the other hand, the electronic occupation $n_B$ shows weak temperature dependence,
as shown in Fig. \ref{condfU}-(b).
The conductance maximum are approximately at $n_B=0.5$ and $1.5$, which coincides with the condition of most probable sequential
tunneling: $\mu=\epsilon_{1,2}$ and $\mu+\epsilon_{1,2}=\epsilon_8$.

\subsection{Coupled-quantum dots}

When the coupling strength $t_H$ between the dots is finite, the system now represents double quantum dots and
all 16 many-body molecular states take part in the transport.
As the first example, we consider the case of $U \rightarrow \infty$ and $U_I=J=0$.
Then, low-lying states are
 $\epsilon_0=0$,
 $\epsilon_{1,2}=\epsilon_d-t_H$,
 $\epsilon_{3,4}=\epsilon_d+t_H$,
and $\epsilon_{5,6,7,8}=2\epsilon_d$ from  Table \ref{table1}.
Due to the large Coulomb potential $U$, one can see that the double occupation on each quantum dot is prohibited.
And one expects that sequential tunneling occurs dominantly for two conditions of $\epsilon_d=t_H$ and $\epsilon_d=-t_H$.
The former corresponds to the transition between $\mid 0\rangle$ and $\mid 1,2\rangle$, and the latter is that between
$\mid 1,2\rangle$ and $\mid 5,6,7,8\rangle$.

\begin{figure}
\centering
\includegraphics[width=0.3\textwidth]{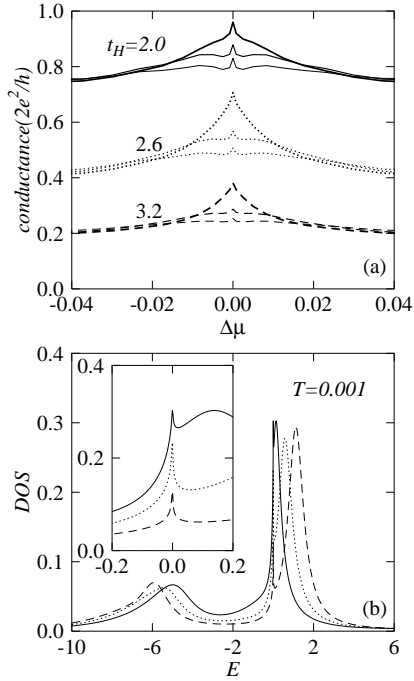}
\caption{
In (a), the conductance (in unit of $2e^2/h$) of the coupled-quantum dot with $U=\infty$ and $U_I=0$ are shown
as a function of the chemical potential difference for temperatures $T=0.001$ (thick), $0.005$ (medium), and
$0.01$ (thin), respectively. In (b), we plot the density of states at $\Delta \mu=0.0$ and $T=0.001$.
The value of the interdot interaction parameter for each of the curves is
$t_H=2.0$ (solid), $2.6$ (dotted), and $3.2$ (dashed) in both panels
and $W=10.0$, $\epsilon_d=-2.5$, and $\mu_0=0$.
\label{DcondIU}
}
\end{figure}

In Fig. \ref{DcondIU}, we examine the conductance as a function of the chemical potential difference
in the vicinity of the latter case.
For a given $\epsilon_d=-2.5$, we compare calculated conductance for $t_H=2.0$, $2.6$, and $3.2$.
Among three cases, overall conductance for $t_H=2.0$ shows the largest value.
It is interesting because the largest one will be the case $t_H=2.6$
according to the sequential tunneling condition of $\epsilon_d=-t_H$.
We attribute this to the level renormalization owing to the electron correlation.
On the other hand, sharp peaks are found around $\Delta\mu=0$, whose height increases as temperatures are lowered.
The peaks are found to result from the Kondo effect as inferred
from the density of states in Fig. \ref{DcondIU}-(b).

Actually, the similar calculation is already performed in Ref. \cite{Aguado1},
where double peaks of the conductance around $\Delta\mu=0$ differently from the present result are observed.
We attribute the discrepancy between a single peak and a double peak predicted in each work
to the difference in the formulation of the problem.
While in Ref. \cite{Aguado1}, the localized basis such as $c_{\alpha\sigma}^\dagger\mid e\rangle$ is used,
we use the diagonalizing basis shown in Table \ref{table1}.
Strictly speaking, the present work treats the double-dot system as a single-quantum dot
with multi-level molecular states, which leads to the modified coupling strengths
between the dots and the leads, weighted by $\xi^{\alpha\sigma}_{mm'}$
in Eq. (\ref{Convert}).
 Therefore, although both approaches adopt the NCA, the details of the Feynman diagrams are different,
and we expect that the results of both approaches would converge by including more crossing diagrams.

\begin{figure}
\centering
\includegraphics[width=0.3\textwidth]{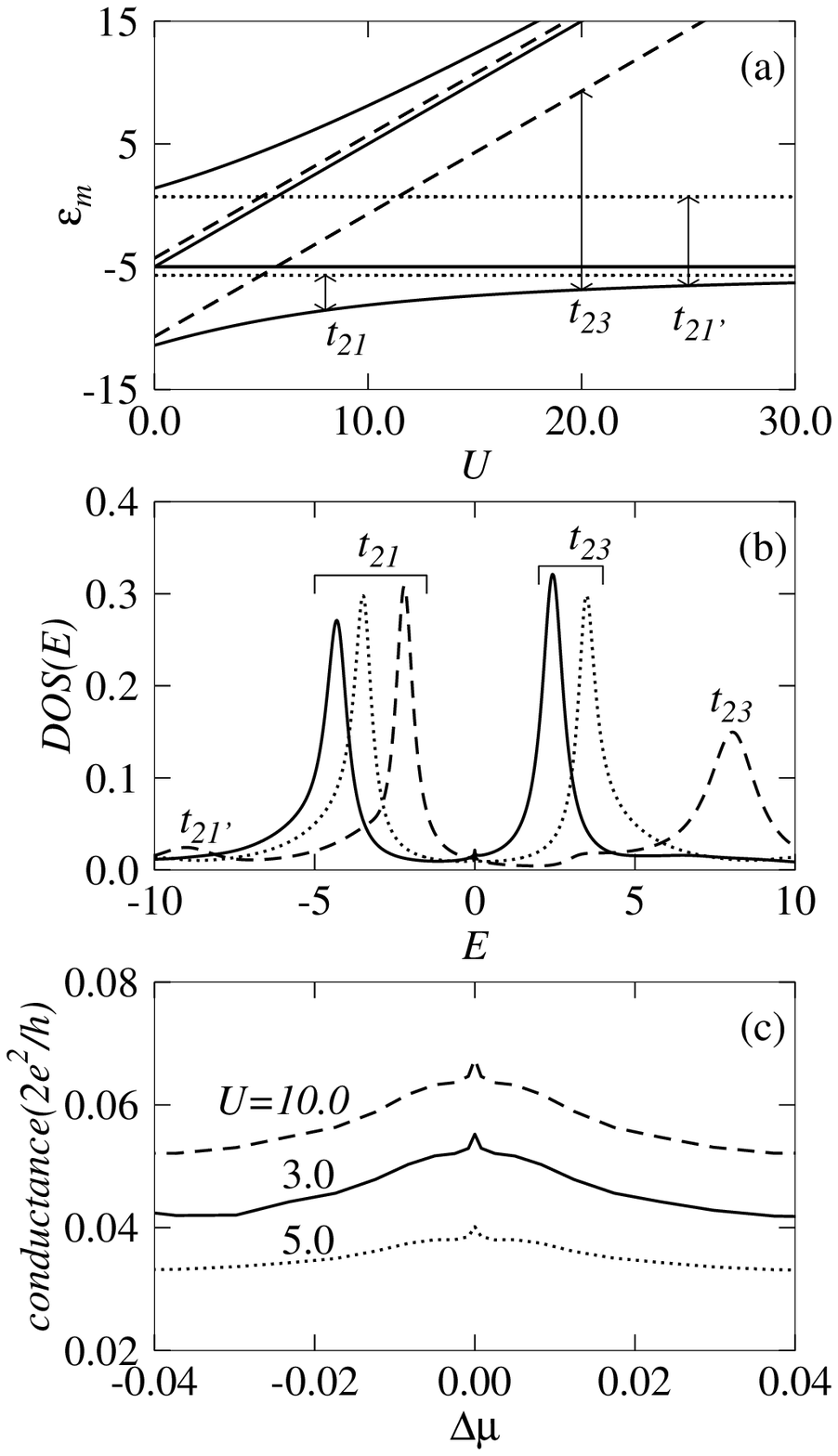}
\caption{
For the coupled quantum dots of Fig. \ref{DcondIU}, the variation of energy levels is plotted in (a)
as a function of Coulomb potential $U$ where one-, two-, three-particle states are represented by solid, dotted, and dashed lines,
respectively.
The arrows indicate possible transitions into one-particle $(t_{21}, t_{21'})$ and three-particle $(t_{23})$ state
from the ground (two-particle single) state.
In (b), we show the density of state for finite Coulomb potentials $U=3.0$ (solid), $5.0$ (dotted), and $10.0$ (dashed), respectively,
at temperature $T=0.003$ and display the transitions corresponding to each peak.
In (c), the conductance (in unit of $2e^2/h$) is shown as a function of the chemical potential difference.
\label{DcondFU}
}
\end{figure}

Finally, we consider the coupled quantum dots with finite Coulomb potential.
When the Coulomb potential becomes comparable to the dot-dot interaction $t_H$, their competition gives
rise to the rich electronic structure and all the energy levels may be relevant to the transport.
In Fig. (\ref{DcondFU})-(a), we show the variation of energy levels
as a function of the Coulomb potential in the case of isolated coupled dots.
As the Coulomb potential decreases from infinity, it is found that
more levels fall into the range of relevant energy.
In other words, this means that various transitions between states become available
and are responsible for more peaks in the density of states as shown in Fig. (\ref{DcondFU})-(b).
Due to the detailed change of energy levels,
the conductance is also found to be largely modified.
In Fig. (\ref{DcondFU})-(c), the conductance are shown for three different Coulomb potentials.
Compared to that of the infinite Coulomb potential case in Fig. (\ref{DcondIU})-(c),
calculated results are largely suppressed.
This is because the transition energies determined from the competition of various interactions are too large
for electrons to tunnel through dots, which is similar to the Coulomb blockade effect for large potential $U$.
Nevertheless, one can see sharp peaks in the calculated conductance at $\Delta\mu=0$.
These peaks result from the Kondo effect as in the case of the infinite Coulomb potential,
meaning that correlated transport still occurs even in small Coulomb potential.
We find that the heights of the conductance at $\Delta\mu=0$ are much larger than those of the master equation approach,
however, smaller than those of NRG (not shown here).\cite{Lee,Izumida}
This means that our approach accounts for correlated behavior of electrons partially.

\section{Summary}

In summary, we formulate the electron transport through two laterally coupled quantum dots by extending the auxiliary operator method to a multi-level case,
and derive the non-equilibrium Green's function in a conserving way.
By using the generating functional technique,
we present exact expressions for the current through the system, as well as the densities of states,
occupancies, and spin correlations of the dots.
To obtain the Luttinger-Ward functional, we include the first-order diagram (non-crossing approximation).
For the validity of our results, we examine various situations and compare calculated results
with those of previous NCA and exact NRG approaches.
We find that our formulation encompasses the previous NCA results successfully, however
gives the deviated behavior from the NRG method.
This means that the present method accounts for the correlated behavior partially and
the vertex correction is needed for more accurate description of transport.
Nevertheless, since the present theory cope with all the ranges of the Coulomb energy and occupancies
as well as time-dependent voltages, it can be applied to reveal transport properties of various double-dot problems.

\acknowledgments{
Numerical calculations are performed on the supercomputer, BlueFern, at the University of Canterbury, New Zealand.
}

\appendix
\section{\label{dcomm}commutation relations}

In this appendix we show that the auxiliary particle representation,
\begin{eqnarray}
c^\dagger_{\alpha\sigma}=\sum_{mm^\prime}\xi^{\alpha\sigma}_{m,m^\prime} d_m^\dagger d_{m^\prime}
\label{ccom}
\end{eqnarray}
gives the correct commutation relations
$[c_{\alpha\sigma},c^\dagger_{\alpha^\prime\sigma^\prime}]_+=
\delta_{\alpha\alpha^\prime} \delta_{\sigma\sigma^\prime}$.
Here, a quasi-particle operator $d_m$ is fermionic (bosonic) when
the number of particles in a state it represents is odd (even), and
is assumed to satisfy the commutation relation $[d_m,d^\dagger_{m^\prime}]_\pm=\delta_{mm^\prime}$.
It is important to note that the expansion coefficient
$\xi^{\alpha\sigma}_{mm^\prime}=\langle m\!\mid c^\dagger_{\alpha\sigma}\mid\! m^\prime\rangle$ is nonzero
only if the number of particles in $\mid m\rangle$ is larger than that in $\mid m^\prime\rangle$ by one.
This means that Eq. (\ref{ccom}) is the combination of fermion and boson operators.

Now, we calculate the commutation relation,
\begin{eqnarray}
[c_{\alpha\sigma},c^\dagger_{\alpha^\prime\sigma^\prime}]_+=
\sum_{m_1,m_2,m_3,m_4}
\xi^{\alpha\sigma *}_{m_2 m_1}
\xi^{\alpha^\prime\sigma^\prime  }_{m_3 m_4}[d^\dagger_{m_2}d_{m_1}, d^\dagger_{m_3}d_{m_4}]_+.
\label{comm1}
\end{eqnarray}
To calculate the right-handed side, it is convenient to separate the sums into the bosonic $(B)$ and fermionic $(F)$ terms,
\begin{eqnarray}
\sum_{m_1,m_2,m_3,m_4}
&=& \sum_{m_2,m_3} \left( \sum_{m_1\in F} +\sum_{m_1\in B}\right )
\left( \sum_{m_4\in F} +\sum_{m_4\in B}\right ) \nonumber\\
&=&
 \left( \sum_{m_1\in F,m_2\in B} +\sum_{m_1\in B,m_2\in F}\right )
 \left( \sum_{m_4\in F,m_3\in B} +\sum_{m_4\in B,m_3\in F}\right ) \nonumber\\
&=&
 \left( \sum_{m_1,m_3\in F}\sum_{m_2,m_4\in B} +\sum_{m_1,m_3\in B}\sum_{m_2,m_4\in F}\right ).
\label{class}
\end{eqnarray}
Here, in the second line we use the fact that $c_{\alpha\sigma}$ and $c_{\alpha\sigma}^\dagger$ are the products of fermionic and
bosonic operators, and in the third line we display the collection of non-zero terms.
By substituting Eq. (\ref{class}) into Eq. (\ref{comm1}), and using $[d_m,d^\dagger_{m^\prime}]_\pm=\delta_{mm^\prime}$,
we arrive at,
\begin{eqnarray}
[c_{\alpha\sigma},c^\dagger_{\alpha^\prime\sigma^\prime}]_+ =
\sum_{m m^\prime m^{\prime\prime}}\left (
 \xi^{\alpha\sigma *}_{m^{\prime\prime} m} \xi^{\alpha^\prime\sigma^\prime}_{m^{\prime\prime} m^\prime}
+\xi^{\alpha\sigma *}_{m^\prime m^{\prime\prime} } \xi^{\alpha^\prime\sigma^\prime}_{m m^{\prime\prime}}
\right ) d^\dagger_m d_{m^\prime}.
\end{eqnarray}
Furthermore, since the expansion coefficients satisfy the orthogonality relation,
\begin{eqnarray}
\sum_{m^{\prime\prime}}\left (
 \xi^{\alpha\sigma *}_{m^{\prime\prime} m} \xi^{\alpha^\prime\sigma^\prime}_{m^{\prime\prime} m^\prime}
+\xi^{\alpha\sigma *}_{m^\prime m^{\prime\prime} } \xi^{\alpha^\prime\sigma^\prime}_{m m^{\prime\prime}}
\right ) =
\langle m\mid
[c_{\alpha\sigma},c^\dagger_{\alpha^\prime\sigma^\prime}]_+\mid m^\prime\rangle =
\delta_{m m^\prime} \delta_{\alpha\alpha^\prime} \delta_{\sigma\sigma^\prime},
\end{eqnarray}
the commutation relation is simplified to
\begin{eqnarray}
[c_{\alpha\sigma},c^\dagger_{\alpha^\prime\sigma^\prime}]_+ =
\delta_{\alpha\alpha^\prime} \delta_{\sigma\sigma^\prime} \sum_m d_m^\dagger d_m
=\delta_{\alpha\alpha^\prime} \delta_{\sigma\sigma^\prime} Q.
\end{eqnarray}
Thus, in the subspace $Q=1$, the combination of quasi-particle operators leads to the correct commutation relation
between $c_{\alpha\sigma}$ and $c^\dagger_{\alpha^\prime\sigma^\prime}$.

\section{\label{PQt}Expressions for physical quantities}

In this section we derive the exact expressions for observables
in terms of the Green's functions.
As examples, we show the procedure for evaluating the current,
\begin{eqnarray}
I_{\alpha\sigma}(t) = q\frac{d}{dt}\left\langle\sum_{k}a_{k\alpha\sigma}^\dagger(t)
a_{k\alpha\sigma}(t)\right\rangle_{GC}
=\frac{iq}{\hbar}\sum_k\left\langle
T^{\alpha*}_{k\sigma} c_{\alpha\sigma}^\dagger(t) a_{k\alpha\sigma}(t)
-T^\alpha_{k\sigma} a^{\dagger}_{k\alpha\sigma}(t)c_{\alpha\sigma}(t) \right\rangle_{GC},
\end{eqnarray}
Green's functions,
\begin{eqnarray}
{\cal G}^R_{\alpha\sigma}(t,t') &=& \frac{1}{i\hbar}\theta(t-t')
\langle [c_{\alpha\sigma}(t),c^\dagger_{\alpha\sigma}(t')]_+\rangle_{GC},  \nonumber\\
\end{eqnarray}
and the occupation number of electrons in each quantum dot,
\begin{eqnarray}
\langle n_{\alpha\sigma}(t)\rangle &=& \left\langle c_{\alpha\sigma}^\dagger(t) c_{\alpha\sigma}(t) \right\rangle_{GC}.
\end{eqnarray}
To do this, we attach fictitious field
$e^{-ip_{\alpha\sigma}(\tau)}$ to $T^{\alpha}_{k\sigma}$ for current, and
add fictitious energy $h_{\alpha\sigma}(t)$ to $\epsilon_{\alpha\sigma}$ for the average number.
Then, from the generating functional $\cal W$ the above quantities can be calculated as,
\begin{eqnarray}
I_{\alpha\sigma}(t) &=&-\left. iq\frac{\delta\cal W}{\delta \Delta p_{\alpha\sigma}(t)}
\right|_{p_{\alpha\sigma}=h_{\alpha\sigma}=0}, \nonumber\\
{\cal G}_{\alpha\sigma}(\tau,\tau') &=& \left.
\frac{\delta [{\cal W}-{\cal W}^{(0)}]}{\delta g_{\alpha\sigma}(\tau,\tau')}
\right|_{p_{\alpha\sigma}=h_{\alpha\sigma}=0}, \nonumber\\
\langle n_{\sigma\alpha}(t)\rangle  &=&-i\hbar \left. \frac{\delta\cal W}{\delta \Delta h_{m}(t)}
\right|_{p_{\alpha\sigma}=h_{\alpha\sigma}=0},
\end{eqnarray}
where $p_{\alpha\sigma}(t)\!=\! \pm\Delta p_{\alpha\sigma}(t)/2$ and
$h_{\alpha\sigma}(t)\!=\!\pm\Delta h_{\alpha\sigma}(t)/2$
are assumed on the upper $(+)$ and lower $(-)$ branches of the Keldysh contour.\cite{Negele,Oh}
With the fictitious fields, the evaluation of the generating functional is straightforward
because bare Green's function is simply changed as,
\begin{eqnarray}
g_{\alpha\sigma}(\tau,\tau^\prime) &\rightarrow&
e^{ip_{\alpha\sigma}(\tau)}g_{\alpha\sigma}(\tau,\tau^\prime)e^{-ip_{\alpha\sigma}(\tau^\prime)}\nonumber\\
g_m^{-1}(\tau,\tau^\prime) &\rightarrow &
(i\hbar\partial_\tau\!-\!\epsilon_m[\epsilon_{\alpha\sigma}\!+\!h_{\alpha\sigma}(\tau)]-\!\lambda\!)
\delta(\tau\!-\!\tau^\prime).
\end{eqnarray}

In order to evaluate the functional derivatives, we expand the generating functional in series,
${\cal W} = \sum_{n=0}^{\infty} {\cal W}^{(n)}$ where
\begin{eqnarray}
{\cal W}^{(0)} &=& \sum_{p}\varsigma_p {\rm Tr}\ln[ g_p^{-1}/i\hbar],\nonumber\\
{\cal W}^{(n)} &=& -\sum_{p}\frac{\varsigma_p}{n}
\oint
g_{p}(\tau,\tau^\prime) \tilde \Sigma^{(n)}_p(\tau^\prime,\tau)d\tau d\tau^\prime.
\end{eqnarray}
Here, $g_p(\tau,\tau^\prime)$ contains the fictitious fields, and
$\tilde \Sigma^{(n)}_p(\tau^\prime,\tau)$ represent all the proper and improper  $n$-th order self-energies.
By performing the functional derivatives we obtain,
\begin{eqnarray}
\frac{\delta{\cal W}}{\delta \Delta h_{\alpha\sigma}(t)} &=&
-\sum_m\varsigma_m\frac{\partial\epsilon_m}{\partial\epsilon_{\alpha\sigma}}
\oint
\frac{\delta h_{\alpha\sigma}(\tau)}{\delta \Delta h_{\alpha\sigma}(t)}\left[
g_{m}(\tau,\tau^\prime)+
g_{m}(\tau,\tau_1)\tilde\Sigma_{m}(\tau_1,\tau_2)
g_{m}(\tau_2,\tau^\prime)
\right] d\tau d\tau^\prime,\nonumber\\
\frac{\delta{\cal W}}{\delta \Delta p_{\alpha\sigma}(t)} &=& i\oint
\frac{\delta p_{\alpha\sigma}(\tau)}{\delta \Delta p_{\alpha\sigma}(t)}\left[
g_{\alpha\sigma}(\tau,\tau^\prime)\tilde\Sigma_{\alpha\sigma}(\tau^\prime,\tau)
-\tilde\Sigma_{\alpha\sigma}(\tau,\tau^\prime) g_{\alpha\sigma}(\tau^\prime,\tau)
\right] d\tau d\tau^\prime.
\end{eqnarray}
By expressing
$\tilde\Sigma= \sum_n\tilde\Sigma^{(n)} = \Sigma+ \Sigma g \Sigma +
\Sigma g \Sigma g \Sigma+\ldots=g^{-1}G\Sigma = \Sigma G g^{-1}$
with proper self-energy $\Sigma$, and performing the Keldysh rotation for the projection onto the real time,
we finally obtain \begin{eqnarray}
I_{\alpha\sigma}(t) &=&
q \Re \int_{-\infty}^{\infty} dt^\prime
\left[
 G^K_{\alpha\sigma}(t,t^\prime)\Sigma_{\alpha\sigma}^A(t^\prime,t)
+G^R_{\alpha\sigma}(t,t^\prime)\Sigma_{\alpha\sigma}^K(t^\prime,t)
\right], \nonumber\\
{\cal G}_{\alpha\sigma}(\tau,\tau') &=& \Sigma_{\alpha\sigma}(\tau,\tau')+
\oint \Sigma_{\alpha\sigma}(\tau,\tau_1)G_{\alpha\sigma}(\tau_1,\tau_2)\Sigma_{\alpha\sigma}(\tau_2,\tau')
d\tau_1 d\tau_2, \nonumber\\
\langle n_{\alpha\sigma}(t)\rangle &=& i\hbar  \sum_m\varsigma_m
\left(\frac{\partial \epsilon_m }{\partial \epsilon_{\alpha\sigma}} \right)
G^<_m(t,t).
\end{eqnarray}
For static cases, since Green's functions depend only on the difference between the time arguments,
the expression for the current is further reduced in energy representation of Eq. (\ref{Fourier}),
\begin{eqnarray}
I_{\alpha\sigma} &=&
\frac{q}{2\pi\hbar} \Re \int_{-\infty}^{\infty} dE
\left[
 G^{>}_{\alpha\sigma}(E)\Sigma_{\alpha\sigma}^{<}(E)
-G^{<}_{\alpha\sigma}(E)\Sigma_{\alpha\sigma}^{>}(E) \right], \nonumber\\
&=&
\frac{q}{2\pi\hbar} \Re \int_{-\infty}^{\infty} dE
\left[
 g^{>}_{\alpha\sigma}(E)\Sigma_{\alpha\sigma}^{<}(E)
-g^{<}_{\alpha\sigma}(E)\Sigma_{\alpha\sigma}^{>}(E) \right],
\end{eqnarray}
where in the second line we make use of Eq. (\ref{Glessas}).

\section{\label{Icon}Current conservation}

The current conservation can be shown by concentrating on one of $n$-th order diagrams
in the generating function, which consist of $n$ conduction or $2n$ auxiliary particle Green's functions.
Each diagram can be expressed either in terms of $n$ conduction Green's functions, or in terms of
$2n$ auxiliary particle Green's functions. Since they represent the same diagram, we can write
\begin{eqnarray}
\sum_{\alpha\sigma}\frac{1}{n}
\oint
g_{\alpha\sigma}(\tau,\tau^\prime) \tilde \Sigma^{(n)}_{\alpha\sigma}(\tau^\prime,\tau)d\tau d\tau^\prime
=
-\sum_{m}\frac{\varsigma_m}{2n}
\oint
g_{m}(\tau,\tau^\prime) \tilde \Sigma^{(n)}_m(\tau^\prime,\tau)d\tau d\tau^\prime.
\end{eqnarray}
By summing all diagrams in the generating functional, we arrive at
\begin{eqnarray}
\sum_{\alpha\sigma} \oint
G_{\alpha\sigma}(\tau,\tau^\prime) \Sigma_{\alpha\sigma}(\tau^\prime,\tau) d\tau^\prime
=-\sum_{m}\oint\frac{\varsigma_m}{2}
G_{m}(\tau,\tau^\prime) \Sigma_m(\tau^\prime,\tau) d\tau^\prime.
\end{eqnarray}
Additionally, by applying the Keldysh rotation onto real time, the above relation becomes
\begin{eqnarray}
\sum_{\alpha\sigma}\int dt' \left[ G_{\alpha\sigma}^K(t,t')\Sigma_{\alpha\sigma}^A(t',t)
+ G_{\alpha\sigma}^R(t,t')\Sigma^K_{\alpha\sigma}(t',t)\right]
=\nonumber\\
-\sum_{m}\frac{\varsigma_m}{2}\int dt' \left[ G^K_m(t,t')\Sigma^A_m(t',t) + G^R_m(t,t')\Sigma^K_m(t',t)\right].
\end{eqnarray}
Using this relation, the sum of currents through both tunneling barriers can be written as
\begin{eqnarray}
\sum_{\alpha\sigma} I_{\alpha\sigma}(t) &=&
-q\Re \sum_{m}\frac{\varsigma_m}{2}\int dt' \left[ G^>_m(t,t')\Sigma^<_m(t',t) - G^<_m(t,t')\Sigma^>_m(t',t)
\right. \nonumber\\
&&~~~~~~~~~~~~~~~ \left. +G^{++}_m(t,t')\Sigma^{++}_m(t',t) -
\Sigma^{++}_m(t,t') G^{++}_m(t',t) \right].\nonumber
\end{eqnarray}
In static case, we obtain, by using the energy representation,
\begin{eqnarray}
\sum_{\alpha\sigma} I_{\alpha\sigma} =
-\frac{q}{2\pi\hbar} \Re \sum_{m}\frac{\varsigma_m}{2}\int dE \left[ G^>_m(E)\Sigma^<_m(E) - G^<_m(E)\Sigma^>_m(E)
\right] =0, \nonumber
\end{eqnarray}
where Eq. (\ref{Gmless}) is used.



\begin{thebibliography}{25}
\bibitem{Wiel} W. G. van der Wiel, S. De Franceschi, J. M. Elzerman, T. Fujisawa, S. Tarucha and L. P. Kouwenhoven,
 Rev.\ Mod.\ Phys.\ {\bf 75}, 1 (2003); R. Hanson, L. P. Kouwenhoven, J. R. Petta, S. Tarucha, L. M. K. Vandersypen,  Rev.\ Mod.\ Phys.\ {\bf 79}, 1217 (2007).  
\bibitem{Ng} T. K. Ng and P. A. Lee, Phys.\ Rev.\ Lett.\ {\bf 61}, 1768 (1988); L. I. Glazman and M. E. Raikh, JETP\ Lett.\ {\bf 47} 452 (1988). 
\bibitem{Goldhaber} D. Goldhaber-Gorden, H. Shtrikman, D. Mahalu, D. Abusch-Magder, U. Meirav, and M. A. Kastner, Nature\  {\bf 391}, 156 (1998).  
\bibitem{Schmid} J. Schmid, J. Weis, K. Eberl, and K. v. Klitzing, Phys.\ Rev.\ Lett.\ {\bf 84}, 5824 (2000); 
H. Jeong, A. M. Chang, and M. R. Melloch, Science\ {\bf 293}, 2221 (2001);  
U. Wilhelm, J. Schmid, J. Weis, and K. v. Klitzing, Physica\ E\ {\bf 14}, 385 (2002); 
A. W. Holleitner, R. H. Blick, A. K. H\"{u}ttel, K. Eberl, and J. P. Kotthaus, Science\ {\bf 297}, 70 (2002). 
\bibitem{Loss} D. Loss and D. P. DiVincenzo, Phys.\ Rev.\ A\ {\bf 57}, 120 (1998). 
\bibitem{Jones} B. A. Jones, C. M. Varma, and J. W. Wilkins, Phys.\ Rev.\ Lett.\ {\bf 61}, 125 (1988).      
\bibitem{Georges}A. Georges and Y. Meir,  Phys.\ Rev.\ Lett.\ {\bf 82}, 3508 (1999).                                         
\bibitem{Izumida}W. Izumida and O. Sakai, Phys.\ Rev.\ B\ {\bf 62}, 10260 (2000).  
\bibitem{Aguado}R. Aguado and D. C. Langreth,  Phys.\ Rev.\ Lett.\ {\bf 85}, 1946 (2000).                                    
\bibitem{Mravlje}J. Mravlje, A. Ram\u{s}ak, and T. Rejec, Phys.\ Rev.\ B\ {\bf 73}, 241305(R) (2006).                 
\bibitem{Lee}M. Lee, M. Choi, R. L\'{o}pez, R. Aguado, J. Martinek, and R. \u{Z}itko, cond-mat.mes-hall/0911.0959.  
\bibitem{Aguado1} R. Aguado and D. C. Langreth,  Phys.\ Rev.\ B\ {\bf 67}, 245307 (2003).         
\bibitem{Petta} J. R. Petta, A. C. Johnson, J. M. Taylor, E. A. Laird, A. Yacoby, M. D. Lukin,
C. M. Marcus, M. P. Hanson, A. C. Gossard, Science\ {\bf 309}, 2180 (2005). 
\bibitem{Koppens} F. H. L. Koppens, C. Buizert, K. J. Tielrooij, I. T. Vink, K. C. Nowack, T. Meunier, L. P. Kouwenhoven,
and L. M. K. Vandersypen, Nature\ {\bf 442}, 766 (2006). 
\bibitem{Johnson} A. C. Johnson, J. R. Petta, C. M. Marcus, M. P. Hanson, and A. C. Gossard, Phys.\ Rev.\ B\ {\bf 72}, 165308 (2005).         
\bibitem{Ono} K. Ono, D. G. Austing, Y. Tokura, and S. Taruchar, Science {\bf 297}, 1313 (2002).  
\bibitem{Wingreen} N. S. Wingreen and Y. Meir, Phys.\ Rev.\ B\ {\bf 49}, 11040 (1994).  
\bibitem{Costi} T. A. Costi, J. Kroha, and P. W{\"o}lfle, Phys.\ Rev.\ B\ {\bf 53}, 1850 (1996).  
\bibitem{Haule} K. Haule, S. Kirchner, J. Kroha, and P. W{\"o}lfle, Phys.\ Rev.\ B\ {\bf 64}, 155111 (2001).  
\bibitem{Mahan} G. D. Mahan, in {\it Many-particle physics}, 2nd ed. (Plenum, New York, 1990).  
\bibitem{NZA} Z. Zout and P. W. Anderson, Phys.\ Rev.\ B\ {\bf 37}, 627 (1988).
\bibitem{Negele} J. W. Negele and H. Orland, in {\it Quantum Many-particle Systems}, (Addison-Wesley, 1988); 
Y. Ustumi, H. Imamura, M. Hayashi, and H. Ebisawa, Phys.\ Rev.\ B\ {\bf 66}, 024513 (2002). 
\bibitem{Oh} J. H. Oh, D. Ahn, and S. W. Hwang, Phys.\ Rev.\ B\ {\bf 72}, 165348 (2005). 
\bibitem{Luttinger} J. M. Luttinger and J. C. Ward,   Phys.\ Rev.\ {\bf 118}, 1417 (1960).
\bibitem{Coleman} P. Coleman, Phys.\ Rev.\ B\ {\bf 29}, 3035 (1984).  
\end{thebibliography}
\end{document}